\documentclass[%
reprint,
superscriptaddress,
amsmath,amssymb,
aps,
prb
floatfix,
showpacs,
]{revtex4-2}

\usepackage{graphicx}
\usepackage{dcolumn}
\usepackage{bm}
\usepackage{hyperref}
\usepackage{ulem}

\usepackage[english]{babel}
\usepackage[utf8]{inputenc}
\usepackage{amssymb,amsfonts,amsmath,mathtools,mathrsfs}
\usepackage{relsize}
\usepackage{mdframed}
\usepackage[margin=0.6in]{geometry}
\usepackage{enumitem}
\usepackage{graphicx}
\usepackage{float}
\usepackage{url,hyperref}
\usepackage[table,usenames,dvipsnames]{xcolor}
\usepackage{gensymb}
\usepackage{grffile}

\hypersetup{
	colorlinks=true,
	linkcolor=Blue,
	filecolor=magenta,      
	urlcolor=Blue,
	citecolor=Blue,
}
\usepackage[bottom]{footmisc}
\usepackage[capitalize]{cleveref}
\usepackage{textcomp}
\usepackage{braket} 

\setcitestyle{line}

\usepackage{amsmath,amssymb,wasysym}

\makeatletter
\def\@fnsymbol#1{\ensuremath{\ifcase#1\or *\or \dagger\or \ddagger\or
   \mathsection\or \mathparagraph\or \|\or **\or \dagger\dagger
   \or \ddagger\ddagger \else\@ctrerr\fi}}
    \makeatother

\begin{document}

\title{Crystal Growth and anisotropic magneto-transport properties of semimetallic LaNiSb\textsubscript{3} }

\address{Department of Chemistry, Indian Institute of Technology Delhi, New Delhi 110016, India}
\author{Haribrahma Singh}
\address{Department of Chemistry, Indian Institute of Technology Delhi, New Delhi 110016, India}

\author{Aarti Gautam}
\address{Department of Chemistry, Indian Institute of Technology Delhi, New Delhi 110016, India}
\author{Prabuddha Kant Mishra}
\address{Department of Chemistry, Indian Institute of Technology Delhi, New Delhi 110016, India}
\author{Rie Y. Umetsu}
\address{Institute for Materials Research, Tohoku University, 2-1-1 Katahira, Aoba-ku, Sendai 980-8577, Japan}
\address{Center for Science and Innovation in Spintronics, Tohoku University, 2-1-1 Katahira, Aoba-ku, Sendai 980-8577, Japan}

\author{Ashok Kumar Ganguli}\email[E-mail: ]{ashok@chemistry.iitd.ac.in}
\address{Department of Chemistry, Indian Institute of Technology Delhi, New Delhi 110016, India}
\address{Department  of Chemical Sciences, Indian Institute of Science Education and Research, Berhampur, Odisha 760003, India}

\begin{abstract}

Single crystals of LaNiSb$_3$ were grown using the Sn-flux method. Structural characterization confirms that LaNiSb$_3$ crystallizes in the orthorhombic $Pbcm$ space group with lattice parameters $a = 13.0970(2)$ Å, $b = 6.1400(4)$ Å, and $c = 12.1270(4)$ Å. Electrical resistivity measurements demonstrate metallic behavior over the entire temperature range of 3–300 K. The magnetoresistance exhibits a positive anisotropic response, attaining a maximum of~8\% for $H \parallel b$, with a pronounced crossover from quadratic to nearly linear field dependence. Angular-dependent MR measurements reveal a pronounced twofold symmetry upon magnetic field rotation within both the $ab$ and $ac$ crystallographic planes up to 50 K, indicating anisotropic charge transport. Hall resistivity measurements show predominantly electron-type conduction at high temperatures, with an increasing hole contribution upon cooling. The multiband character is further corroborated by the violation of Kohler’s scaling and is well described within a semiclassical two-band framework.  Collectively, these results suggest LaNiSb$_3$ has anisotropic multiband electronic transport and could be compulsive candidate to explore structure-property correlation as a topological semimetal.

\end{abstract}


\maketitle

\section{Introduction}

Materials that host square-net pnictogen layers have emerged as versatile platform for exploring unconventional electronic states driven by symmetry protection. Compounds featuring nonsymmorphic crystal lattices are of particular interest, as glide-plane and screw-axis symmetries can enforce band degeneracies and produce protected crossings that remain robust even in the presence of strong spin–orbit coupling  \cite{Young2015,Liang2017,Klemenz2019}. Such symmetry-enforced constraints have been shown to generate Dirac-like dispersions, hourglass fermions, and nodal-line features in a range of layered pnictides, motivating the search for new materials in which nonsymmorphic operations and orbital topology cooperate to produce electronic nontriviality.

A central structural motif in LnTSb$_3$  (Ln $=$ Lanthanides, T $=$ Transition metal) family is the Sb square net, a nearly two-dimensional p-orbital network capable of hosting linearly dispersing bands due to its delocalized electronic character and weak interlayer coupling \cite{Klemenz2019,schoop2016,Hidetoshi2016}. In well-studied analogues such as ZrSiS-type compounds, these square nets give rise to high-mobility carriers, small Fermi surfaces, and large nonsaturating magnetoresistance—properties often associated with semimetallic band structures shaped by symmetry-protected crossings \cite{Topp2017,Neupane2016}. When square nets coexist with nonsymmorphic lattice symmetry, the resulting band topology may exhibit additional constraints, as symmetry operations limit allowed hybridization and can protect specific nodal features across the Brillouin zone.

LaNiSb$_3$ is a member of the broader family of isostructural antimonides, LnNiSb$_3$ (Ln = La, Ce, Pr, Nd, etc.), which constitute a chemically tunable platform for exploring the interplay between crystal structure, electronic correlations, and magnetism \cite{Macaluso2004,Thomas2004a,Thomas2007}. In particular, members such as CeNiSb$_3$ and related compounds exhibit rich correlated-electron phenomena, including Kondo-lattice behavior and long-range magnetic ordering \cite{Macaluso2004,Thomas2004a,Thomas2007}. In parallel, transition-metal substitution at the Ni site offers an effective means to tune bandwidth and carrier concentration, thereby modifying the electronic states associated with the Sb square-net sublattice \cite{Yang2022,Rai2022}. Together, these features establish the LnNiSb$_3$ series as a versatile framework for exploring structure–property relationships in square-net–based systems.

Closely related compounds in the LnCrSb$_3$ family, particularly LaCrSb$_3$, have attracted significant attention due to their pronounced structural anisotropy, coexistence of localized and itinerant magnetism, and indications of Berry-curvature–driven anomalous Hall effects \cite{Kumar2021a}. These findings highlight the importance of lattice symmetry and electronic topology in shaping emergent transport responses in rare-earth antimonides. In this context, LaNiSb$_3$, crystallizing in the orthorhombic $Pbcm$ space group with Sb square nets embedded in a nonsymmorphic lattice, provides an opportunity to examine analogous physics in a Ni-based system where strong magnetic order is absent \cite{Paul2025}.

Despite this favorable structural setting, LaNiSb$_3$ remains comparatively underexplored \cite{Thomas2007}. 
Given that nonsymmorphic symmetry can enforce band degeneracies and that square-net motifs frequently host dispersions conducive to Dirac-like carriers, a systematic investigation of its transport properties is warranted. Here, we present a comprehensive study of LaNiSb$_3$ through electrical resistivity, magnetotransport, and Hall-effect measurements. By analyzing its anisotropic magnetoresponse and multiband transport characteristics, we assess its electronic structure and place LaNiSb$_3$ within the broader class of square-net–based materials exhibiting unconventional transport phenomena.


\begin{figure}[t!]
\includegraphics[width=1
\columnwidth,angle=0,clip=true]{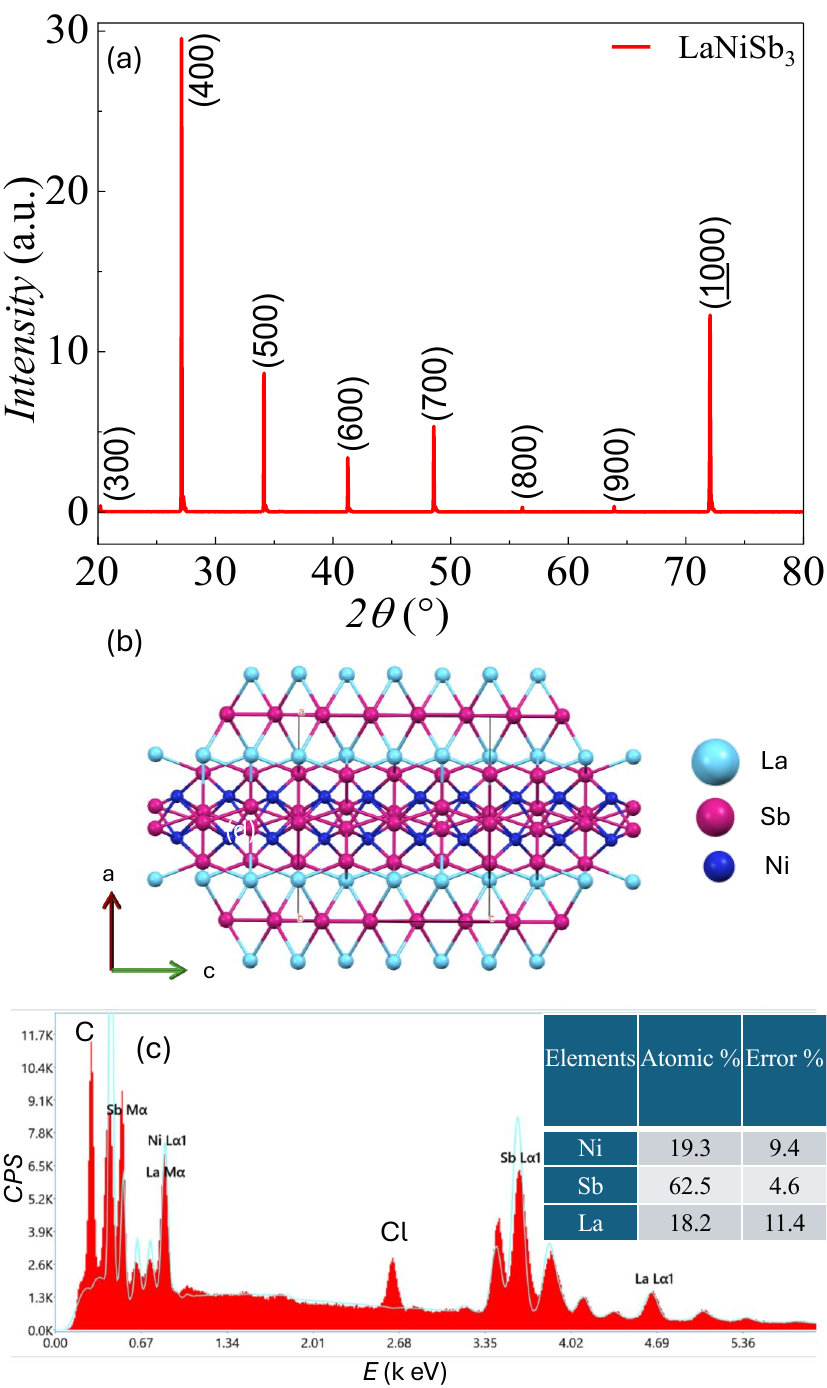}
\caption{(a) Powder X-ray diffraction pattern of LaNiSb$_3$ single crystal. (b) Crystal structure of LaNiSb$_3$ (c) Energy-dispersive X-ray (EDX) elemental spectrum of LaNiSb$_3$.}
\label{Fig. 1}
\end{figure}

\section{Experimental Details}

Single crystals of LaNiSb$_3$ were grown using a conventional solid-state synthesis method. Elemental La, Ni, and Sb were combined in a molar ratio of 1:1:3 and mixed with Sn flux in a 1:1 sample-to-flux proportion. The mixture was sealed under high vacuum (10$^{-5}$ bar) in a quartz ampoule, heated to 1000 $\celsius$ for 48 h, and then slowly cooled to room temperature at a rate of 10 °C per hour. After the reaction, shiny black plate-like crystals were isolated by dissolving the excess flux in hydrochloric acid.

Room-temperature single-crystal X-ray diffraction (SCXRD) measurements were performed on a Bruker D8 Quest diffractometer equipped with a microfocus Mo K$_\alpha$ radiation source ($\lambda$ = 0.71073 \r{A}). Data processing was carried out using the APEX4 suite \cite{Apex4}, and absorption effects were corrected via the multi-scan method implemented in SADABS \cite{SADABS}. The crystal structure was solved with SHELXS97 \cite{Sheldrick2008} and refined using SHELXL through the OLEX2 platform. 
Elemental composition was examined on a single crystal using a TESCAN MAGNA FESEM system equipped with EDX. Minor C K$\alpha$ and Cl K$\alpha$ signals in the spectra originate from the carbon tape and residual HCl employed during crystal extraction.
Magnetization measurements were conducted using a SQUID magnetometer (MPMS3, Quantum Design). Electrical transport properties were measured using a standard four-probe configuration in a Quantum Design Physical Property Measurement System (PPMS) over 2-300 K and in magnetic fields up to 9~T.

\section{Results and discussion}

\begin{table}[t!]
    
\scriptsize\addtolength{\tabcolsep}{-0.5pt}
\caption{\label{table1}Single Crystal Structural information of LaNiSb$_3$.}
\begin{ruledtabular}
\begin{tabular}{lc} 
Empirical formula & LaNiSb$_3$ \\  
Formula weight & 562.87\\
Space group & $Pbcm$ \\ 
Crystal system & Orthorhombic\\
{\it a}(\r{A}) & 13.0552(6) \\
{\it b}(\r{A}) & 6.1418(3) \\
{\it c}(\r{A}) & 12.1256(5) \\
Volume(\r{A}$^3$) & 972.26(8) \\
Z & 8 \\
Temperature(K) &	300.00 \\
$\rho_{cal}$ (g/cm$^3$)  &  7.691\\
Radiation 	& MoK$\alpha$ ($\lambda$ = 0.71073 \r{A})\\
2$\theta$ range for data collection & 	6.24 to 28.31 \\
Index ranges 	& -17 $\leq$ h $\leq$ 17, -8 $\leq$ k $\leq$ 6, -16 $\leq$ l$\leq$ 14\\
Reflections collected 	& 37388 \\ 
Independent reflections 	&  1269 [R$_{int}$ = 0.0782]\\
Data/restraints/parameters 	& 1269/0/25 \\
Goodness-of-fit on F$^2$ 	&  0.993 \\
Final R indexes [I$\geq$2$\sigma$ (I)] & 	R$_1$ = 0.0395, wR$_2$ = 0.1203 \\
Final R indexes [all data] 	& R$_1$ = 0.0400, wR$_2$ = 0.1210\\
Largest diff. peak/hole ( e {A$^{-3}$}) 	& 3.08/-3.23 \\
\end{tabular}
\end{ruledtabular}
\end{table}

\begin{table}[t!]
\scriptsize\addtolength{\tabcolsep}{-1pt}
\scriptsize
\caption{\label{table2}Fractional Atomic Coordinates and Equivalent Isotropic Displacement Parameters (\AA$^2$). $U_\text{eq}$ is defined as $1/3$ of the trace of the orthogonalised $U_{ij}$ tensor.}
\centering
\begin{ruledtabular}
\begin{tabular}{l c c c c c r}
Atom & Wyckoff Site & Occu. & $x$ & $y$ & $z$ & $U_\text{eq}$ \\
\hline
La1 & 4c & 1 & 0.8000(5) & 3/4 & 1/2 & 0.007(2) \\
La2 & 4d & 1 & 0.8045(5) & 0.2631(9) & 1/4 & 0.007(5) \\
Sb1 & 4c & 1 & 0.7158(1) & 1/4 & 0.0001 & 0.008(2) \\
Sb2 & 4d & 1 & 0.7102(2) & 0.7469(5) & 1/4 & 0.008(2) \\
Sb3 & 4c & 1 & 0.5245(2) & 3/4 & 0.0000 & 0.011(2) \\
Sb4 & 4d & 1 & 0.4464(0) & 0.6151(5) & 1/4 & 0.008(2) \\
Sb5 & 8e & 1 & 0.9985(5) & 0.4918(3) & 0.3761(5) & 0.008(1) \\
Ni1 & 8e & 1 & 0.6021(2) & 0.4686(2) & 0.1361(4) & 0.009(2) \\ 
\end{tabular}
\end{ruledtabular}
\end{table}

\subsection{Crystal structure.}

LaNiSb$_3$ crystallizes in an orthorhombic structure with the $Pbcm$ space group (lattice parameters: $a$=13.0970(2)\AA, $b$=6.1400(4)\AA, $c$=12.1270(4)\AA) \cite{Thomas2007}. The structure is characterized by layered arrangements that host nearly square or rectangular two-dimensional atom nets. Within this framework, two crystallographically inequivalent La sites (La1 and La2) are present. Their coordination environments differ: La1 forms a square antiprismatic polyhedron, whereas La2 adopts a monocapped square antiprismatic geometry. The La–La separation along the $a$-axis is relatively large, approximately 8.32 \AA, while the corresponding distance within the bc-plane is significantly shorter, around 4.25 \AA. The refined crystal structure of LaNiSb$_3$ obtained in this work is consistent with that reported by Thomas et al. \cite{Thomas2007}. The crystallographic orientation of the synthesized single crystals was validated using the face-indexing technique.

\begin{figure}[t!]
\begin{center}
    
\includegraphics[width=1. \columnwidth,angle=0,clip=true]{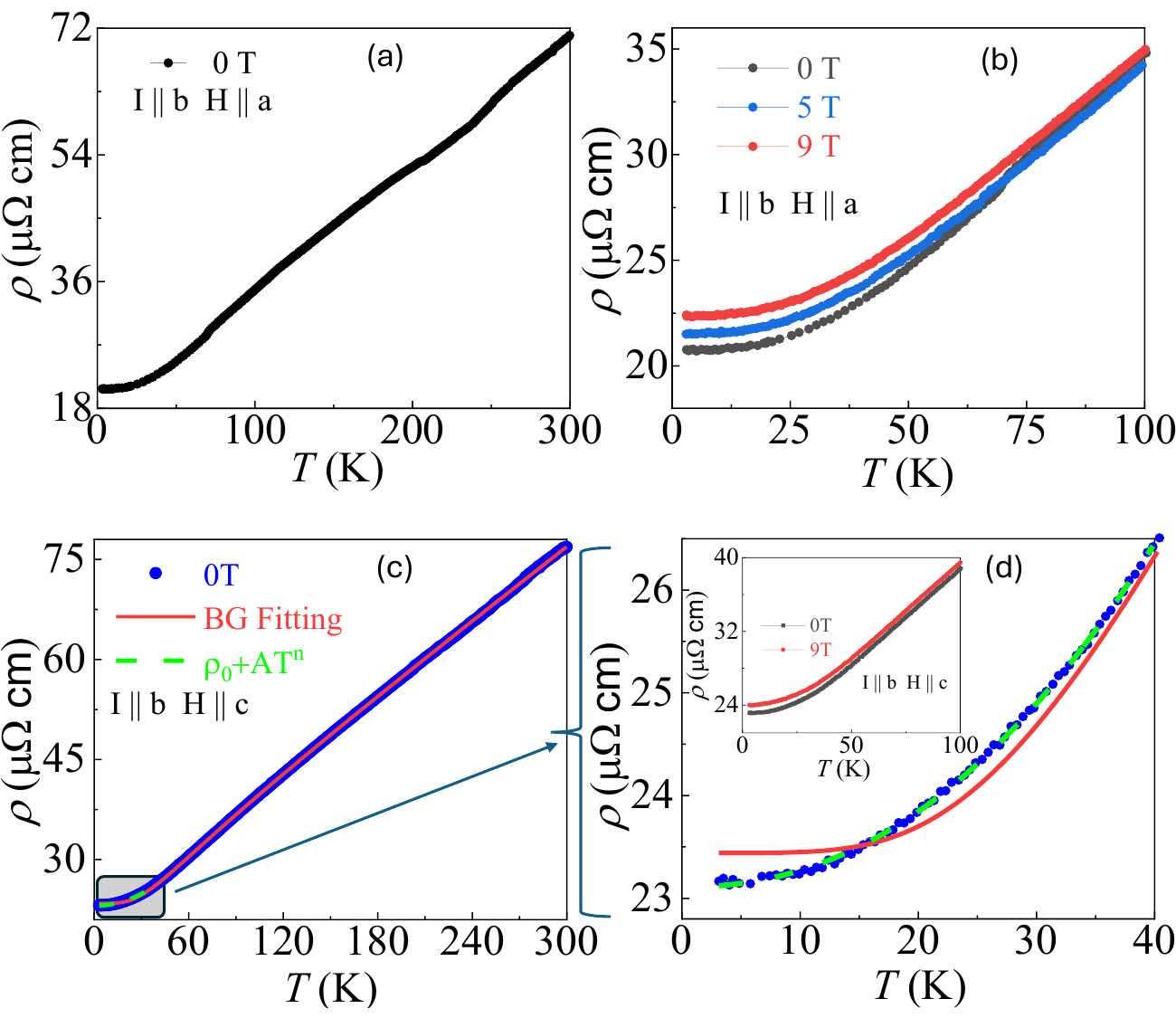}
\caption{Temperature dependence of the electrical resistivity, $\rho(T)$, for current applied along the $b$ axis. 
(a) Zero-field resistivity for the configuration $I \parallel b$. 
(b) $\rho(T)$ measured under applied magnetic fields of 0, 5, and 9~T with $H \parallel a$. 
(c) Bloch--Gr\"uneisen (BG) fitting of the resistivity data for $I \parallel b$ and $H \parallel c$ in the temperature range 3--300~K. 
(d) Comparison of the resistivity fits using the power-law relation $\rho(T)=\rho(0)+AT^{\alpha}$ and the BG model. 
The inset highlights the $\rho(T)$ behavior at higher magnetic fields.}

\label{Fig. 2}
\end{center}
\end{figure}

\subsection{Transport properties}

\begin{figure*}
\begin{center}
    
\includegraphics[width=2. \columnwidth,angle=0,clip=true]{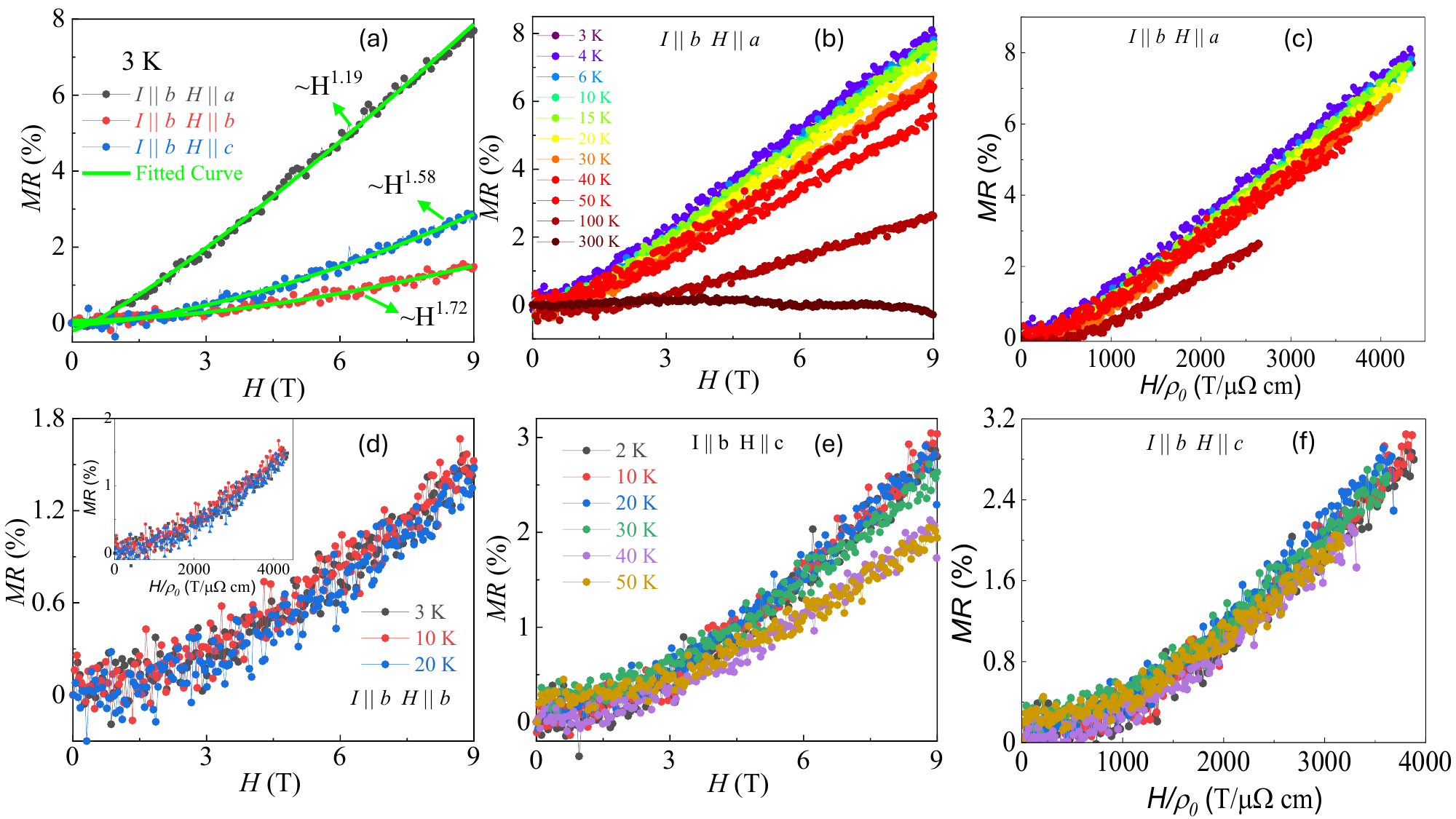}
\caption{Magnetoresistance (MR) under different field--current configurations.
(a) MR as a function of magnetic field $H$ at 3~K for $I \parallel b, H \parallel a$; 
$I \parallel b, H \parallel b$; and $I \parallel b, H \parallel c$. 
The solid green line represents linear fit for the $H \parallel a$ configuration. 
The MR follows a power-law dependence $\sim H^{n}$ with the fitted exponents indicated in the panel.
(b) MR versus $H$ for $I \parallel b, H \parallel a$ measured from 3~K to 300~K.
(c) Kohler’s plot of MR as a function of $H/\rho_{0}$ for $I \parallel b, H \parallel a$.
(d)  MR versus $H$ for $I \parallel b, H \parallel b$ at selected temperatures (3, 10, and 20~K), inset shows corresponding Kohler's plot. 
(e) MR versus $H$ for $I \parallel b, H \parallel c$ at temperatures between 2~K and 50~K.
(f) Kohler’s plot of MR versus $H/\rho_{0}$ for $I \parallel b, H \parallel c$.}

\label{Fig. 3}
\end{center}
\end{figure*}

\begin{figure*}
\includegraphics[width=1.5\columnwidth,angle=0,clip=true]{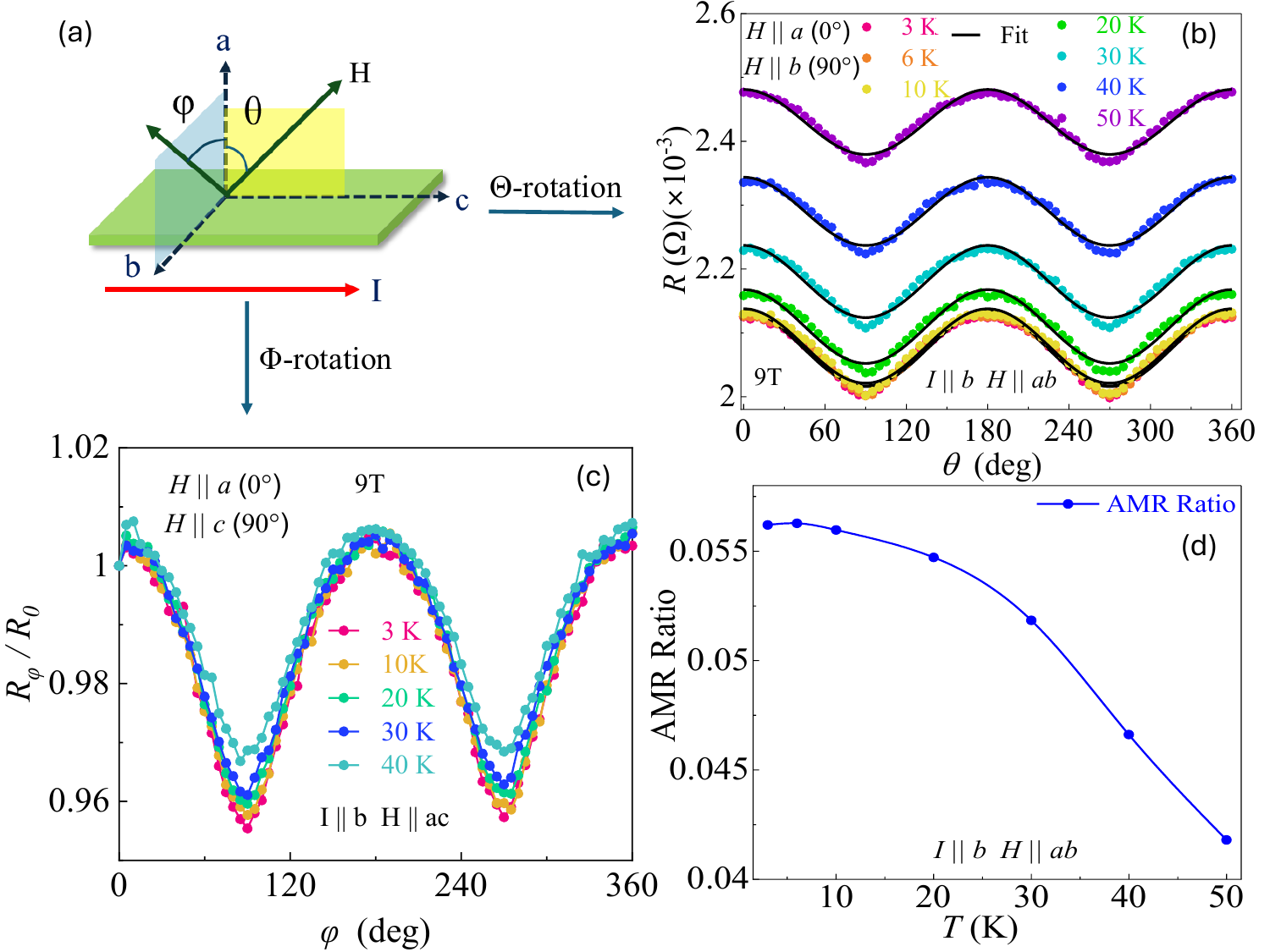}
\caption{Angular magnetoresistance:
(a) Schematic illustration of the measurement geometry. The electric current is applied along the $b$ axis ($I \parallel b$). 
The polar angle $\theta$ is defined for rotation of the magnetic field in the $ab$ plane ($H \parallel ab$), 
while $\phi$ corresponds to rotation in the $ac$ plane ($H \parallel ac$).
(b) Angular dependence of resistance $R(\theta)$ measured at 9~T for $I \parallel b$, $H \parallel ab$ at different temperatures (3--50~K). 
Solid lines represent fits to the data with equation 3.
(c) Normalized resistivity $R_\phi/R_0$ as a function of $\phi$ at 9~T for $I \parallel b$, $H \parallel ac$ at various temperatures. 
(d) Temperature dependence of the anisotropic magnetoresistance (AMR) ratio extracted for $I \parallel b$, $H \parallel ab$.}
\label{fig:4}
\end{figure*}

The resistivity ($\rho$) was measured as a function of temperature from 300~K to 3~K using a four-probe technique with the current applied along the b-axis. 
$\rho(T)$ vs $T$ is shown in Fig.~\ref{fig:2}. The residual resistivity ratio (RRR = $\rho_{300~\mathrm{K}}/\rho_{2~\mathrm{K}}$) was obtained to be $\approx 5$ for $H \parallel c$. The $\rho(T)$ exhibits metallic behavior with an almost linear temperature dependence from 300~K to 40~K followed by a crossover to power-law behavior at lower temperatures $T \le 40~\mathrm{K}$. Fig.~\ref{fig:2} represents an enlarged view of the low-temperature region ($T \leq 40~\mathrm{K}$), where the resistivity is well described by.-

$\rho(T) = \rho(0) + AT^\alpha$. Fitting yields an exponent of $\alpha=$2.22. This nearly quadratic temperature dependence indicates that electron–electron scattering dominates transport at low temperatures. In contrast, in the high-temperature region ($T\ge$40 K), the resistivity shows a quasi-linear temperature dependence, characteristic of dominant electron–phonon scattering. This behavior can be quantitatively analyzed using the Bloch–Grüneisen (BG) model \cite{cvijovic2011}.

\begin{equation}
\rho(T)=\rho(T_0)+A\left(\frac{T}{\Theta_R}\right)^5\int_0^{\Theta_R/T}\frac{x^5}{(e^{x}-1)(1-e^{-x})}\,dx,
\label{Bloch}
\end{equation}

where $\rho(T_0)$ (23.42 $\micro \ohm$ cm) is the residual resistivity at $T=$2~K, $A$  is a pre-factor related to the electron-phonon coupling strength, and $\Theta_R$ is the Debye temperature estimated from the resistivity data. From the fit (see \hyperref[Fig_3]{Fig.3(c)}), we obtain $A$= 0.01225(4)$\mu\ohm$cm and $\Theta_R$=177.0(6)K.

Further, $\rho(T)$ was also measured in various applied magnetic fields, below 100~K. Considering the current in the b-axis, the magnetic field can be applied perpendicularly along the a-axis (\hyperref[Fig_2]{Fig. 2(b)}) and c-axis (inset of \hyperref[Fig_2]{Fig. 2(d)}). The enhanced resistivity with field indicates positive magnetoresistance over  the measured temperature range for both configurations. 


\begin{figure*}[ht]
\includegraphics[width= 2\columnwidth,angle=0,clip=true]{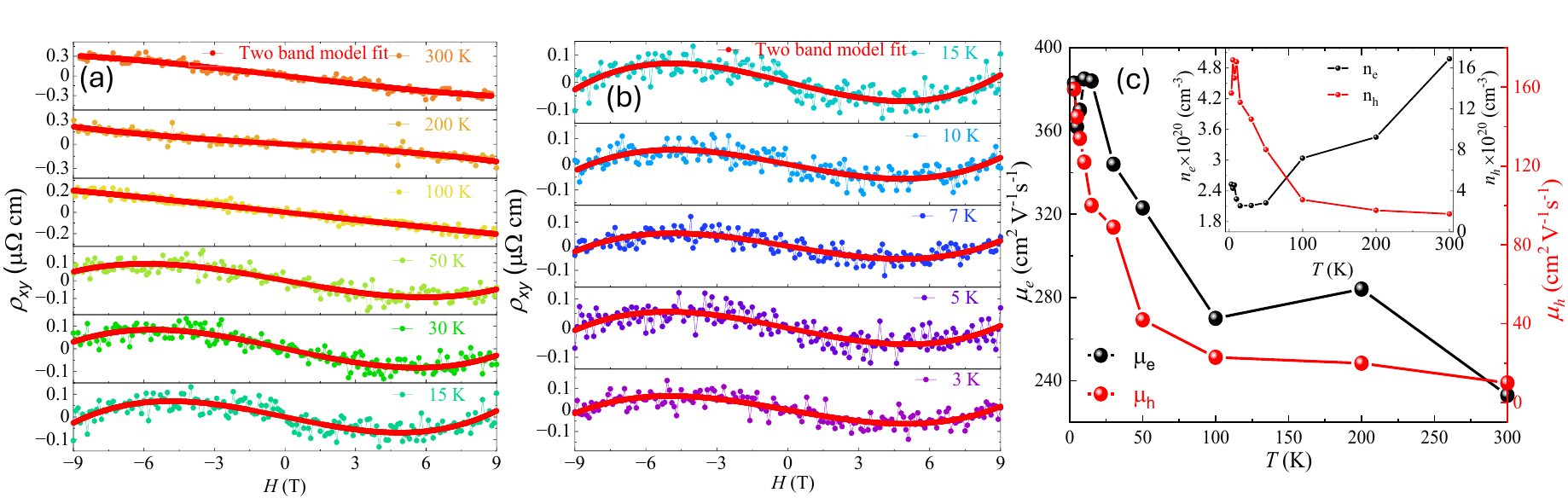}
\caption{Hall effect analysis and two-band model fitting: Field dependence of Hall resistivity $\rho_{xy}$ at (a) Higher temperatures (30--300~K). Solid lines represent fits using the semiclassical two-band model. (b) $\rho_{xy}(H)$ measured at low temperatures (3--15~K) together with two-band model fits.
(c) Temperature dependence of the carrier mobilities $\mu_e$ (black symbols) and $\mu_h$ (red symbols) extracted from the two-band fitting. Inset: Temperature dependence of the electron and hole carrier densities ($n_e$ and $n_h$).}
\label{Fig. 5}
\end{figure*}

\subsection{Magnetotransport properties}


The field dependence of the magnetoresistance (MR), measured for various magnetic-field configurations, is presented in \hyperref[Fig_3]{Fig. 3}. The magnetoresistance is calculated through the relation MR\% = $[\rho(H)-\rho(0)]/\rho(0)\times 100$. In this relation, the $\rho(H)$ and $\rho(0)$ are the resistivity in the applied field and in the absence of field. The MR data at $T=3$~K are shown in \hyperref[Fig_3]{Fig. 3(a)} for a field applied along all three crystal axes. 
Magnetotransport measurements reveal a pronounced positive magnetoresistance (MR) at low temperatures with a clear anisotropic dependence on the magnetic-field orientation. The largest MR magnitude is observed for $H \parallel a$, where the magnetic field is applied perpendicular to the Sb square net lying in the $bc$ plane. In contrast, the MR magnitude decreases when the field is oriented parallel to the $bc$ plane. A similar anisotropic MR behavior has been reported in LaAgSb$_2$ \cite{Akiba2022}, suggesting that this feature may be intrinsic to Sb-square net layered compounds.\cite{Akiba2022}. Notably, MR exhibits an approximately linear field dependence in the high-field regime. However, in the low-field region, a clear deviation from linearity is observed, characterized by a crossover from the conventional semiclassical quadratic ($\propto H^2$) dependence at low fields to a linear field dependence at higher fields.
To quantitatively analyze the field dependence, the MR data at 3 K were fitted using a single power-law expression, $\mathrm{MR} \propto H^{n}$, along the three principal crystallographic directions. As shown in \hyperref[Fig_3]{Fig. 3(a)}, the MR follows an single power-law behavior with exponents $n \approx 1.19$ for $H \parallel a$, $n \approx 1.58$ for $H \parallel c$, and $n \approx 1.7$ for $H \parallel b$. The direction-dependent power-law exponents are further associated with the anisotropic Fermi surface topology and carrier dynamics associated with the layered electronic structure \cite{Xu2025,flessa2023}.

The observed non-quadratic field dependence, particularly the quasi-linear MR behavior for $H \parallel a$, indicates a clear deviation from the conventional semiclassical quadratic ($n = 2$) response. Such crossover from quadratic to linear MR has often been associated with Dirac-like linear band dispersion near the Fermi level and, more broadly, with the presence of nontrivial fermionic states \cite{Malick2025, Gautam2026}. Similar quasi-linear MR behavior has been reported in related layered compounds such as LaAgSb$_2$, LaSbTe \cite{Singha2017}, and other materials \cite{Xu2015a,Singha2017,Singh2024,Wang2012}. Nevertheless, further experimental investigations are required to conclusively establish the microscopic origin of the observed behavior in the present system.



Nonuniform scattering rates over the Fermi surface or their dependence on current orientation lead to anisotropic transport behavior, which is often manifested as a violation of Kohler’s rule, implying disparate temperature scaling of scattering rates across the Fermi surface \cite{Xu2015a,Singha2017,Singh2024}. 
\begin{equation}
    MR = f\left(\frac{H}{\rho(0)}\right)
\end{equation}

The MR curves measured at different temperatures can be collapsed onto a single scaling curve. According to Kohler’s law within the semiclassical two-band framework, such scaling is expected when a single characteristic scattering rate governs the transport. However, a slight deviation from this scaling is observed in \hyperref[Fig_3]{Fig. 3(c and f)} and the inset of \hyperref[Fig_3]{Fig. 3(d)} at temperatures above 10~K, which becomes increasingly pronounced at temperatures exceeding 100~K. This deviation suggests the involvement of multiple electron and hole bands with unequal carrier densities and mobilities.  
 
The angular magnetoresistance (AMR) was measured at various temperatures by rotating the sample in a constant magnetic field of 9~T. The measurements were performed for rotations within the ab-plane ($\phi$) and the ac-plane ($\theta$). In the first configuration ($\theta$), the field rotates from perpendicular to the current to parallel to it (\hyperref[Fig_4]{Fig. 4(b)}), whereas in the second configuration ($\phi$), the field is always perpendicular to the current (\hyperref[Fig_4]{Fig. 4(c)}). The schematic illustration of crystal orientation is shown in the \hyperref[Fig_4]{Fig. 4(a)}. The two-fold nature of the data is observed in both configurations \cite{Tang2025}. The similar nature of AMR in both configurations suggests the least Zeeman effect in the scattering. However, the relative anisotropy is higher in the ab-plane. It is important to note that the nature and magnitude of AMR are least affected by increases in temperature. Furthermore, we have fitted the AMR data using \hyperlink{equation3}{equation 3}.

\begin{equation}
    R(\theta)=R_0 + \Delta R \cos{2\theta}
    \label{equation3}
\end{equation}

Where $\theta$ is the angle between the current and the magnetic field. 
R$_\perp$= $R_0 + \Delta R$, R$_\parallel$= $R_0 - \Delta R$, AMR ratio is defined as $|(R_\perp - R_\parallel)/R_\perp|$, The AMR ratio increases as the temperature decreases and nearly saturates at low temperatures as shown in \hyperref[Fig_4]{Fig. 4(d)}.


\subsection{Hall measurements}

 To gain deeper insight into the charge transport mechanism and the nature of charge carriers in LaNiSb$_3$, $\rho_{xy}(T)$ measurements were performed over a temperature range of 3–300 K under magnetic fields up to $\pm 9$ T with the electrical current applied in the $ab$ plane. As shown in \hyperref[Fig_5]{Fig.~5(a)}. At room temperature, $\rho_{xy}$ varies linearly with the applied magnetic field and exhibits a negative slope, indicating that electrons are the dominant charge carriers. However, upon lowering the temperature below 50 K, a clear deviation from linearity appears in the field dependence of $\rho_{xy}(H)$ as depicted in \hyperref[Fig_5]{Fig.~5(b)}. This nonlinearity suggests the emergence of an additional type of charge carrier i;e multi-band transport. Such behavior can be well described within the framework of the semiclassical two-band model, which is expressed as follows:

\begin{equation}\label{Halltwoband}
\rho_{xy} = \frac{H}{e} \frac{n_h \mu_h^2 - n_e \mu_e^2 + (n_h - n_e) \mu_e^2 \mu_h^2 H^2}{(n_e \mu_e + n_h \mu_h)^2 + (n_h - n_e)^2 \mu_e^2 \mu_h^2 H^2}
\end{equation}

In this model, $n_h$ ($\mu_h$) and $n_e$ ($\mu_e$) represent the carrier density (mobility) of holes and electrons, respectively. A representative fit to the $\rho_{xy}$ data is shown in \hyperref[Fig_5]{Fig. 5(a-b)}. The temperature dependence of the extracted parameters are plotted in \hyperref[Fig_5]{Fig. 5(c)}. From the fitting parameters, the electron and hole densities at 3 K are found to be 2.5$\times$10$^{20}$ and 1.35$\times$10$^{21}$ cm$^{-3}$, respectively. These values are lower than those of conventional metals and are often associated with nodal-line band crossings near the Fermi level. Similar carrier concentrations have been reported in Dirac nodal-line semimetals such as LaSbSe ($10^{19}$–$10^{20}$ cm$^{-3}$) \cite{Pandey2022} and LaMn$_{0.86}$Sb$_2$ ($10^{21}$–$10^{22}$ cm$^{-3}$) \cite{Yang2023}, where clear nodal-line features are observed near the Fermi level. Similarly, the obtained carrier mobilities for electrons and holes at 3 K are 3.8$\times 10^2 $ and 1.5$\times 10^2$ cm$^2$ V$^{-1}$ s$^{-1}$, respectively.  \hyperref[Fig_5]{Figure. 5(c)} shows the temperature dependence of carrier densities for electrons and holes. With increasing temperature, the electron density shows a gradual increase, whereas the hole density decreases. In contrast, the mobilities of both carriers decrease monotonically which can be attributed to enhanced phonon scattering at elevated temperatures. At 3~K, the hole density exceeds the electron density, indicating an uncompensated charge-carrier system. As illustrated in \hyperref[Fig_5]{Fig. 5(c)}, The comparable magnitudes of electron and hole mobilities suggest that both carrier types contribute significantly to the overall transport. This multiband nature is also corroborated by the deviation of the MR curves from Kohler’s scaling.

\section*{Conclusions}

High-quality single crystals of LaNiSb$_3$ were successfully grown, crystallizing in the orthorhombic lattice with the space group $Pbcm$. The structure hosts Sb-square net ($4^4$) embedded within a nonsymmorphic framework, a structural motif often associated with nontrivial band topology. Electrical transport measurements confirm metallic behavior with a residual resistivity ratio (RRR) of approximately 5 with nearly linear temperature dependence of resistivity.
Magnetotransport measurements show positive magnetoresistance at low temperatures, exhibiting anisotropy with respect to field orientation. The magnetoresistance is maximimum for $H \parallel a$ (field perpendicular to the Sb square-net) and least for 
$H \parallel b$ (field parallel to the Sb square-net). Notably, the magnetoresistance for $H \parallel a$ exhibits a quasi-linear field dependence, deviating from the conventional quadratic behavior expected for classical metals. AMR measurements display robust twofold symmetry for both $ab$-plane and $ac$-plane field rotations, indicating anisotropic charge transport.
Hall-effect measurements further confirm multiband transport involving both electron- and hole-type charge carriers. The Hall resistivity is well described by a two-band model, yielding carrier densities $n_e = 2.5\times10^{20}$ and $n_h$ = 1.35$\times$10$^{21}$ cm$^{-3}$, and mobilities $\mu_e = 3.8\times 10^2 $ and $\mu_h$ =1.5$\times 10^2$ cm$^2$ V$^{-1}$ s$^{-1}$ for electrons and holes, respectively. The coexistence of carriers with distinct mobilities supports the observed deviation from classical magnetotransport behavior.
Overall, the observation of quasi-linear magnetoresistance together with pronounced multiband Hall behavior strongly indicates possible topological semimetallic characteristics in LaNiSb$_3$. These findings place it within the broader family of square-net–based nonsymmorphic materials, which are known to host symmetry-protected electronic states and potentially unconventional transport phenomena.


\section*{ACKNOWLEDGMENTS}

The authors acknowledge the Central Research Facility (CRF) and Department of Chemistry, IIT Delhi, for EDX and single-crystal XRD facilities. We thank the MPMS3 facility, Department of Physics, IIT Delhi. We acknowledge CRDAM for the PPMS facility proposal, No. 202502-CR KKE-0508, and young researcher fellowship program in ICC, at IMR, Tohoku University. HBS acknowledges the Council of Scientific $\&$ Industrial Research (CSIR) File No. 09/0086(12695)/2021-EMR-I, India, for fellowship. AKG expresses gratitude to SERB for the financial assistance from Project Number CRG/2022/000178.


\bibliographystyle{apsrev4-2}
\bibliography{LaNiSb3}

\end{document}